\documentclass[12pt,prd,amsfonts,nofootinbib,showpacs]{revtex4}
\usepackage{graphicx, epsfig}
\newcommand{\be}{\begin{equation}}
\newcommand{\ee}{\end{equation}}
\newcommand{\bea}{\begin{eqnarray}}
\newcommand{\eea}{\end{eqnarray}}
\begin{document}
%%%%%%%%%%%%%%%%%%%%%%%%%%%%%%%%%%%%%%%%%%%%%%%%%%%%%%%%%%%%%%%%%%%%%%%%
\title{The width difference of rho vector mesons}

%%%%%%%%%%%%%%%%%%%%%%%%%%%%%%%%%%%%%%%%%%%%%%%%%%%%%%%%%%%%%%%%%%%%%%%%
\author{F. V. Flores-Ba\'ez and G. L\'opez Castro}
\affiliation{Departamento de F\'\i sica, Cinvestav, Apartado Postal
14-740, 07000 M\'exico, D.F. M\'exico}
\author{G. Toledo S\'anchez}
\affiliation{Instituto de F\'\i sica, Universidad Nacional Aut\'onoma de
M\'exico, 04510 M\'exico, D.F. M\'exico}
%%%%%%%%%%%%%%%%%%%%%%%%%%%%%%%%%%%%%%%%%%%%%%%%%%%%%%%%%%%%%%
\begin{abstract}
We compute the difference in decay widths of charged and neutral
$\rho(770)$ vector mesons. The isospin breaking arising from mass
differences of neutral and charged $\pi$ and $\rho$ mesons,
radiative corrections to $\rho \to \pi\pi$, and the $\rho \to
\pi\pi\gamma$ decays are taken into account. It is found that the width
difference $\Delta \Gamma_{\rho}$ is very sensitive ot the isospin
breaking in the $\rho$ meson mass $\Delta m_{\rho}$. This result can
be useful to test the correlations observed between the values of these
parameters extracted from experimental data.

\end{abstract}
%%%%%%%%%%%%%%%%%%%%%%%%%%%%%%%%%%%%%%%%%%%%%%%%%%%%%%%%%%%%%%%%%%%%
\maketitle

%%%%%%%%%%%%%%%%%%%%%%%%%%%%%%%%%%%%%%%%%%%%%%%%%%%%%%%%%%%%%%%%%%%%%%%
\section{Introduction.}
%%%%%%%%%%%%%%%%%%%%%%%%%%%%%%%%%%%%%%%%%%%%%%%%%%%%%%%%%%%%%%%%%%%%%%%
Lowest-lying vector mesons undergo predominantly strong interaction
decays. The masses and decay widths of members of the same isomultiplet
will therefore look very similar \cite{pdg}, with small differences
induced by the breaking of isospin symmetry. 
   The isospin breaking effects in the $\rho$ meson parameters have raised
an interest recently, due to both experimental and theoretical reasons 
\cite{Alemany:1997tn, Bijnens:1996kg, Feuillat:2000ch, Cirigliano:2002pv,
Ghozzi:2003yn}. According
to the PDG \cite{pdg} the weighted averages of available measurements are:
\bea
\Delta m_{\rho} &\equiv & m_{\rho^0}-m_{\rho^{\pm}} = (-0.7\pm 0.8) \
\mbox{\rm
MeV}\ ,\\
\Delta \Gamma_{\rho}&\equiv &  \Gamma_{\rho^0}-\Gamma_{\rho^{\pm}} =
(0.3\pm
1.3)\
\mbox{\rm MeV}\ .
\eea
These results are consistent with the absence of isospin breaking in the
$\rho^0-\rho^{\pm}$ system. Note however that  the scale factors
associated with  the above averages are, respectively, 
1.5 and 1.4 \cite{pdg} which reflects an important spread in the yields
from different experiments.

Some recent theoretical calculations of $\Delta m_{\rho}$
seem to confirm the above result. Using a vector-meson dominance model
to parametrize the $\gamma\rho\rho$ vertex, the authors of Ref.
\cite{Feuillat:2000ch} have obtained $\Delta m_{\rho}=(-0.02\pm 0.02)$
MeV. Also, using $1/N_c$ expansion techniques, the authors of
Ref. \cite{Bijnens:1996kg} have obtained $-0.4\ \mbox{\rm MeV} \leq \Delta
m_{\rho} \leq 0.7\ \mbox{\rm MeV}$. On another hand, it has been found
that the width difference of $\rho$ mesons is of great importance to
understand the current discrepancy between the hadronic vacuum
polarization contributions to the muon anomalous magnetic moment obtained
from $\tau$ decay and $e^+e^-$ annihilation data \cite{Alemany:1997tn, 
Cirigliano:2002pv, Ghozzi:2003yn}.

  In this paper we provide an estimate of $\Delta \Gamma_{\rho}$
 by considering the isospin breaking corrections in the exclusive 
modes that contribute to the decay widths of $\rho^{0,\pm}$
vector mesons. A previous estimate of this effect was done in reference
\cite{Alemany:1997tn} taking into account several sources of isospin
breaking as mass differences and other subleading $\rho$ meson decays.
Their result $\Delta \Gamma_{\rho} \approx (-0.42 \pm 0.59)$ MeV
\cite{Alemany:1997tn} is consistent with the world average given in Eq.
(2). Additional contributions to isospin breaking in $\Delta
\Gamma_{\rho}$, including the radiative corrections to the dominant $\rho
\to \pi\pi$ decays, are considered in this paper.

\section{Sources of isospin breaking}

  At a fundamental level, isospin symmetry is broken by the different
masses of $u$ and $d$ quarks and by the effects of electromagnetic
interactions. At the hadronic level all manifestations of isospin breaking
can be traced back to such fundamental sources. In the absence of isospin
breaking, the $\rho^{0,\pm}$ mesons must have equal masses and decay
widths, thus $\Delta m_{\rho}=\Delta \Gamma_{\rho}=0$.

  The dominant decay modes of $\rho$ mesons that are common to charged
and neutral $\rho$'s are the $\pi\pi$ decay and its radiative mode. The
branching fraction of other  modes contributing only to the $\rho^0$ meson
adds up
to \cite{pdg}:
\be
B^0_{rest}= B(\pi^0\pi^0\gamma)+ B(\eta\gamma)+B(\mu^+\mu^-) + B(e^+e^-) +
B(\pi^+\pi^-\pi^0) \approx 5.3 \times 10^{-4} \ .
\ee
There is also a dipole transition $\rho \to \pi\gamma$ which is common to
$\rho^{\pm,0}$ vector mesons with branching fractions of a few times
$10^{-4}$ \cite{pdg}.  Since the $\rho$ meson widths are of order 150 MeV,
all these 
subleading decay modes  will contribute to the width difference at the
tiny level of:
\be
\Delta \Gamma^{\mbox{\rm sub}}_{\rho} \approx 0.08 \ \mbox{\rm MeV}\ .
\ee
Thus, any sizable difference in the decay widths can only originate from
the dominant decay modes. To be more precise, we will define explicitly
the contributions to the width difference as follows:
\bea
\label{delta} \Delta \Gamma_{\rho} &=& \Gamma(\rho^0 \to
\pi^+\pi^-(\gamma),
\omega\leq
\omega_0)-\Gamma(\rho^+ \to \pi^+\pi^0(\gamma), \omega\leq
\omega_0) \nonumber \\
&& \  \  + \ \Gamma(\rho^0 \to \pi^+\pi^-\gamma, \omega\geq
\omega_0)-\Gamma(\rho^+ \to \pi^+\pi^0\gamma, \omega\geq
\omega_0) + \Delta \Gamma^{\mbox{\rm sub}}_{\rho}\ .
\eea
The first two terms in Eq. (\ref{delta}) denote the $\pi\pi$ decay rates
that include virtual and soft-photon corrections and the next two
terms are related to hard-photons in $\pi\pi\gamma$ decays. In the above
equation, $\omega_0$ is an arbitrary value of the photon energy that
separates the decay rates of soft- and hard-photon bremsstrahlung. It is
expected that the $\omega_0$ dependence will cancel in the sum of the
first (second) and third (fourth) terms of Eq. (\ref{delta}).
 
In the following section we will consider in more detail
the isospin breaking corrections to $\rho \to \pi\pi$ and their radiative
decays. Later, we will evaluate the contributions of such corrections to
the r.h.s. of Eq. (\ref{delta}).

\section{Radiative corrections to $\rho \to \pi\pi$ decays}

  At the lowest order (indicated with superscript $0$), the rates of
 $\rho \to \pi\pi$ decays are given by:
\bea
\Gamma^0_0 &\equiv &\Gamma^0(\rho^0 \to \pi^+\pi^-) =
\frac{g^2_{+-}}{48\pi} m_{\rho^0}
v_0^3\ , \\
\Gamma^0_+ &\equiv &\Gamma^0(\rho^+ \to \pi^+\pi^0) =
\frac{g^2_{+0}}{48\pi} m_{\rho^+}
v_+^3\ ,
\eea
where
\be
v_0 \equiv \sqrt{1-\frac{4m_{\pi^+}^2}{m_{\rho^0}^2}},\ \ \ \ {\rm and} \
\ \ \ \ v_+
\equiv 
\sqrt{\left(1-\frac{(m_{\pi^+}- m_{\pi^0})^2}{m_{\rho^+}^2}\right)
\left(1-\frac{(m_{\pi^+}+m_{\pi^0})^2}{m_{\rho^+}^2}\right)}\ ,
\ee
 are the pion velocities in the rest frame of the $\rho^i$ meson 
and $g_{ij}$ are the $\rho\pi^i\pi^j$ coupling constants such that
$g_{+-}=g_{+0}$ owing to the isospin symmetry of strong interactions. In
the limit of isospin symmetry, the masses of neutral and charged pions
($\rho$'s) are the same, $v_0=v_+$, and consequently the three-level decay
rates are equal ($\Gamma_0^0=\Gamma_+^0$). In the following subsections we
will
consider the different ingredients to get the $O(\alpha)$ radiative
corrections to the decay rates given above.

\subsection{Real soft-photon corrections}

\begin{figure}
\includegraphics[width=10cm]{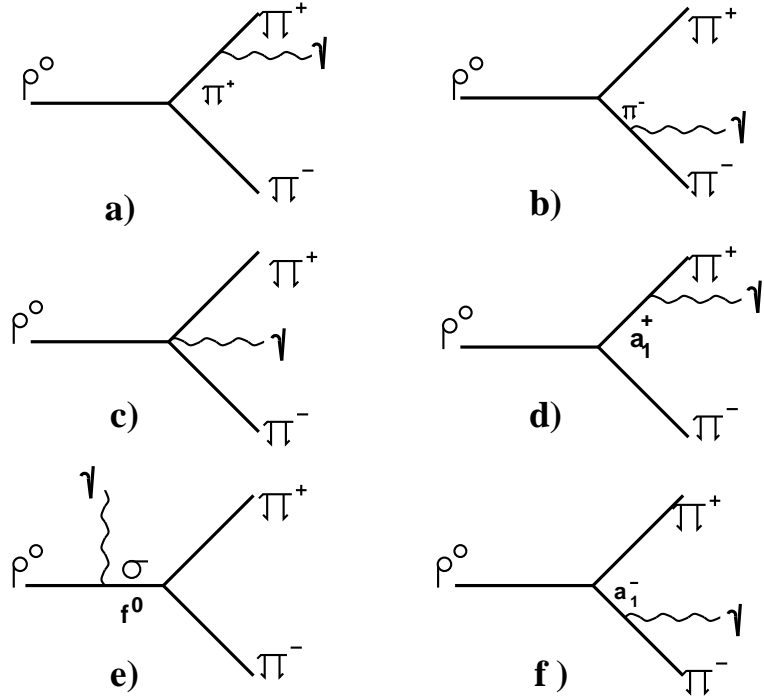}\\
\caption{Feynman graphs for $\rho^0 \to \pi^+\pi^-\gamma$ decays.}
\end{figure}

\begin{figure}
\includegraphics[width=10cm]{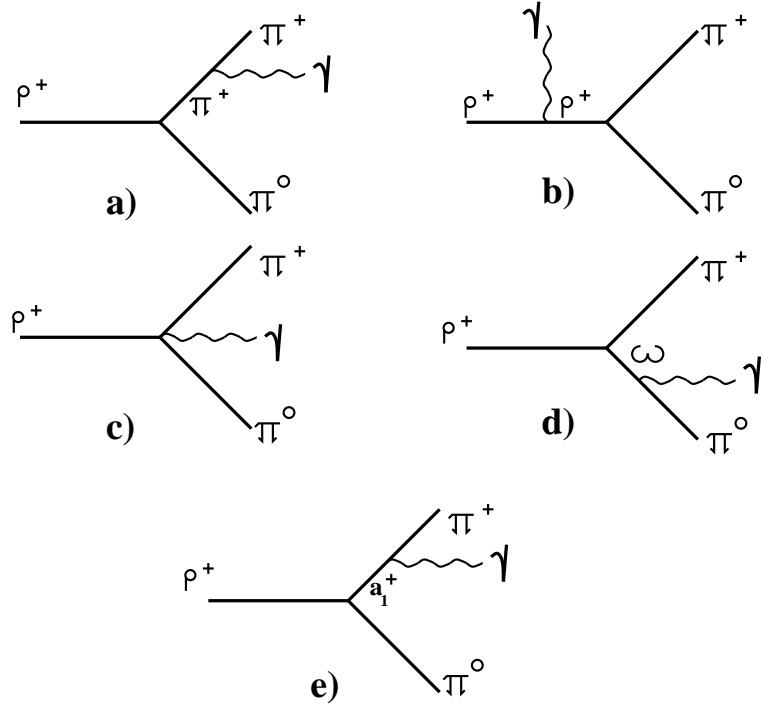}\\
\caption{Feynman graphs for $\rho^+ \to \pi^+\pi^0\gamma$ decays.}
\end{figure}

  In order to get an infrared safe result, the radiative corrections of
order $\alpha$ to Eqs. (6,7) must include the sum of virtual corrections
and soft-photon bremsstrahlung. In order to define the soft-photon
contribution, let us consider the radiative $\rho \to \pi\pi\gamma$ decays
shown in Figures 1 and 2. In each of Figures 1 and 2, the first three
diagrams correspond to the model-independent contributions and the other
graphs denote the model-dependent terms.

The decay amplitude of the  process  $\rho^0(d,\eta)\rightarrow  \
\pi^+(p) \pi^-(p')  \gamma(k,\epsilon^*)$
(four-momenta and polarization four-vectors are indicated 
within parenthesis), see Figure 1, is given by \cite{Singer:1963ae,
Toledo:2007}: 
\bea 
\!\! {\mathcal{M}}(\rho^0 \to \pi^+\pi^-\gamma)\! &=&\! ie g_{+-}\!
\left\{ 
\! (p-p')   \cdot \eta\!
\left(\frac{p'\cdot\epsilon^*}{p'\cdot\epsilon^*} -  
\frac{p\cdot\epsilon^*}{p\cdot\epsilon^*} \right)\! 
+ 2\left[\epsilon^* \cdot \eta- \frac{p \cdot \epsilon^*k\! \cdot \eta}
{p \cdot k} \right]\!\right\}\! + {\mathcal{M}}_{\rm
d,e,f}  
\label{bremssn}
\eea
where the model-independent pieces of the amplitude are given explicitly,
and the remaining terms ${\mathcal{M}}_{\rm d,e,f}$ correspond to the
model-dependent amplitudes arising from Figures
1e)-1f).
Similarly, the decay amplitude for the decay  $\rho^+(d,\eta)\rightarrow \
\pi^+(p) \pi^0(p') \gamma(k,\epsilon^*)$), see Figure 2, is given by the
following expression \cite{Singer:1963ae, Bramon:1993yx, Lopez
Castro:1997dg,
 Lopez Castro:2001qa}:
\bea 
 {\cal M}(\rho^+ \to \pi^+\pi^0\gamma) & = & ieg_{+0} \left \{ \left (
\frac{p.\epsilon^*}{p.k} -
\frac{d.\epsilon^*}{d.k} \right ) (p-p') \cdot \eta
+\left ( \frac{p.\epsilon^*}{p.k} -
\frac{d.\epsilon^*}{d.k} \right ) k\cdot \eta  \right. \nonumber \\
& & \ \ \ \ \ \ \left. + \left[ 2+\frac{\Delta \kappa}{2} \left( 1 +
\frac{\Delta^2_{\pi}}{m^2_{\rho^+}} \right) \right] \left(\frac{d\ \cdot
\epsilon^*}{d \cdot k}k\cdot \eta - \epsilon^* \cdot \eta \right) \right.
\nonumber \\
& & \ \ \ \ \ \ \left. - (2+ \Delta \kappa) \left( \frac{p\ \cdot
\epsilon^*}{p \cdot k}k\cdot \eta - \epsilon^* \cdot \eta \right)
\frac{p\cdot k}{d \cdot k} \right\} + {\cal M}_{\rm d,e} \ ,
\label{bremssc} 
\eea
where $\Delta \kappa$ denotes the anomalous magnetic dipole moment of the
$\rho^+$ vector meson in units of $e/2m_{\rho^+}$ (in our numerical
evaluations we will set $\Delta \kappa =0$), and we have defined
$\Delta^2_{\pi} \equiv m_{\pi^+}^2-m_{\pi^0}^2$.  Once again, only the
model-independent amplitudes for Figures 2a)-2c) have been written
explicitly.

The soft-photon bremsstrahlung is the divergent piece of the amplitudes
(\ref{bremssn},\ref{bremssc}) in the infrared
region ($k \to 0$). The diagrams
contributing to the soft-photon amplitude are shown in Figs. 1a)-1b) for
$\rho^0 \to\pi^+\pi^-\gamma$ and in Figs. 2a)-2b) for $\rho^+ \to 
\pi^+\pi^0\gamma$ decays.  As usual, only photons of energy smaller than
$\omega_0$  must be considered in the real-photon emission rate that is
included in radiative corrections. Thus, the
radiative corrected rates of $\rho^i \to \pi\pi$ decays ($i=0,+$) can be
written in the following general form:
\bea
\Gamma(\rho^i \to \pi\pi(\gamma), \omega\leq\omega_0)&=& \Gamma^0_i +
\Gamma^{\rm 1v}_i + \Gamma^{\rm soft}_i(\omega_0)\  \nonumber \\
&=& \Gamma^0_i \left[ 1+\frac{\Gamma^{\rm 1v}_i}{\Gamma^0_i} +
\frac{\Gamma^{\rm soft}_i(\omega_0)}{\Gamma^0_i}\right] \nonumber \\
&=& \Gamma^0_i \left[ 1+\delta_{\rho^i}\right] \ .
\eea
where $\Gamma^{\rm 1v}_i$ denotes the virtual corrected rate at order
$\alpha$  for $\rho^i \to \pi\pi$ decay and $\Gamma^{\rm
soft}_i(\omega_0)$ is the soft-photon rate of $\rho^i \to \pi\pi\gamma$  
obtained from the first term in Eqs. (\ref{bremssn},\ref{bremssc}).

 The soft-photon rates can be computed in
 analytical form by integrating over photon energies smaller than
$\omega_0$, with the following results, respectively for $\rho^0$ and
$\rho^+$ radiative decays:
\bea
\frac{\Gamma^{\rm soft}_0(\omega_0)}{\Gamma^0_0}&=&
\frac{\alpha}{\pi}\left\{
2\ln\left(\frac{\lambda}{2\omega_0}\right)\left(1+\frac{1+v_0^2}{2v_0}
\ln\left[\frac{1-v_0}{1+v_0}\right]\right)-\frac{1}{v_0}\ln\left(\frac{1-
v_0}{1+v_0}\right) \right. \nonumber\\
&&\left. +\frac{1+v_0^2}{2v_0}\left[\mbox{\rm Li}_2\left(
\frac{1-v_0}{1+v_0}\right)-\mbox{\rm Li}_2\left(\frac{1+v_0}{1-v_0}
\right)+\imath\pi \ln\left( \frac{1-v_0}{1+v_0}\right) \right. \right.
\nonumber \\
 & & \left. \left. 
+\ln\left( \frac{1-v_0}{1+v_0}\right)\ln\left(\frac{4v_0^2}{1-v_0^2}\right)
\right] \right\}\ ,
\label{bssn}
\eea
and
\bea
\frac{\Gamma^{\rm soft}_+(\omega_0)}{\Gamma^0_+}&=&\frac{\alpha}{\pi}
\left\{1 -2\ln2
+ 2\ln\left[\frac{\lambda}{\omega_0}\right]\left(1+
\frac{1}{2v'_+}\ln\left[ \frac{1-v'_+}{1+
v'_+}\right]\right)\right. \nonumber\\ && 
\left. +\frac{1}{2v'_+}\ln\left(\frac{1+v'_+}{1-v'_+}\right)
 -\frac{1}{2v'_+}\left[{\rm Li}_2
\left(\frac{1+v'_+}{1-v'_+}\right)-{\rm Li}_2\left(\frac{1-
v'_+}{1+v'_+}\right) \right. \right. \nonumber \\ &&  
\left. \left. +\ln\left(\frac{v_+^{'2}}{1-v_+^{'2}}  
\right)\ln\left(\frac{1+v'_+}{1-v'_+}\right) + \imath \pi
\ln\left( \frac{1+v'_+}{1-v'_+}\right)\right]\right\}\ ,
\label{bssc}
\eea
where we have defined $v'_+ = v_+/\xi$, with
$\xi=1+\Delta^2_{\pi}/m_{\rho^+}^2$,  and ${\rm Li}(x)$ denotes the Spence
function.

   Observe that the soft-photon rates depend logarithmically upon the
infrared cutoff $\lambda$ (the photon mass regulator) and the photon
energy parameter $\omega_0$. We have checked that Eq. (12) coincides
numerically with the result reported long ago by Cremmer and Gourdin 
\cite{Cremmer:1969er} in the case of $\phi \to K^+K^-\gamma$ decays,
although our
corresponding expressions are written in a different form. Finally, let us
point out that a very small contribution arising from the regular part
(finite when the photon energy goes to zero) of the radiative decay rate
must be added to the r.h.s. of Eq. (11).

\subsection{Virtual corrections}

   The Feynman diagrams of virtual  corrections of $O(\alpha)$ to 
the $\rho \to \pi\pi$ amplitudes are shown in Figures 3 and 4.
As in the real photon case, we will use scalar QED for the
electromagnetic couplings of charged pions and assume that the
$\rho^{\pm}$ electromagnetic vertex (see for example \cite{Bramon:1993yx, 
Lopez Castro:1997dg}) is similar to the $WW\gamma$ gauge boson vertex.  

The sum of virtual corrections to the $\rho^0\to \pi^+\pi^-$ amplitude is
finite
in the ultraviolet region owing to a Ward identity satisfied by the
vertex and self-energy corrections \cite{Cremmer:1969er}. In the case of
the $\rho^+ \to \pi^+\pi^0$ decay, the sum of  virtual corrections  
diverges. One way of getting rid of such divergences is by absorbing
them into 
a redefinition of the strong coupling constant $g_{+0}$.
Instead of dealing with divergent amplitudes, in the
case of $\rho^+\to \pi^+\pi^0$ decays we will follow a procedure
introduced long ago by Yennie \cite{Yennie:63}. According to this
method we pick only the convection terms from virtual corrections
in Figures 4a)-c). The sum of such terms gives rise  to a one-loop
amplitude which is finite in the ultraviolet region  and is also
gauge-invariant and model-independent \cite{Yennie:63}. Furthermore, such
an amplitude contains all the infrared divergent terms of virtual
corrections \cite{Yennie:63}.

\begin{figure}
\includegraphics[width=10cm]{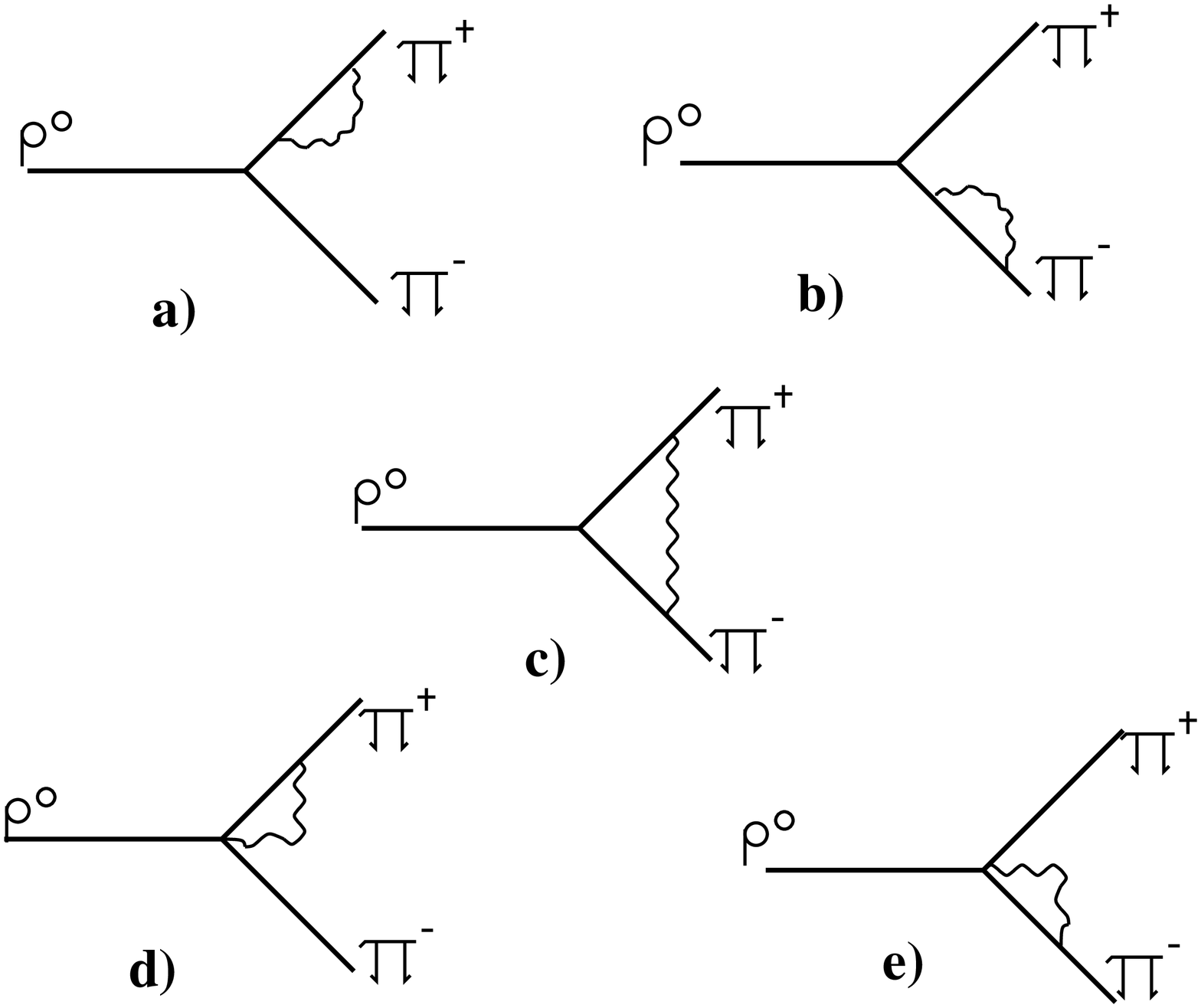}\\
\caption{Feynman graphs for the virtual photonic corrections to
$\rho^0 \to \pi^+\pi^-$ decays.}
\end{figure}

\begin{figure}
\includegraphics[width=10cm]{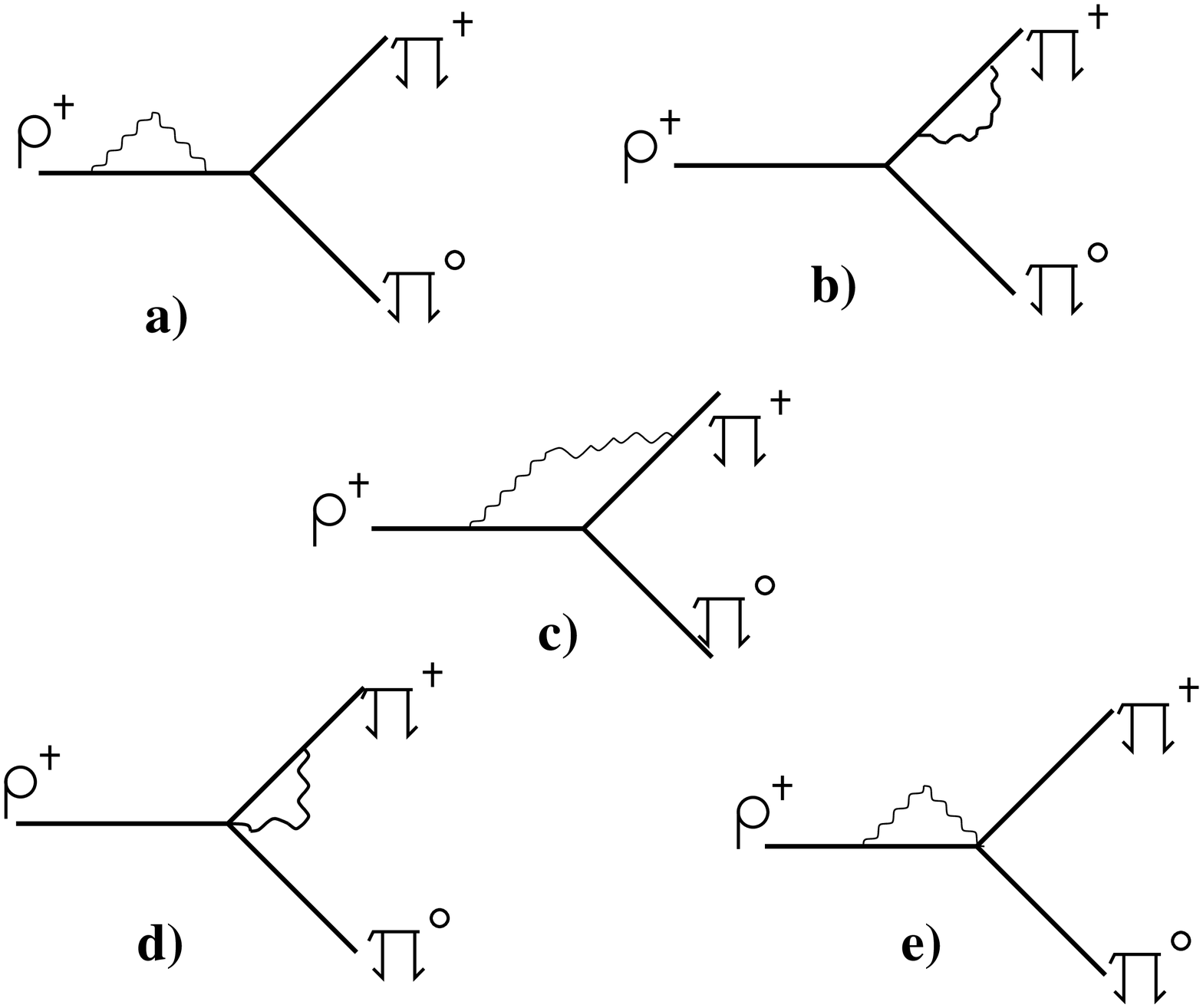}\\
\caption{Feynman graphs for the virtual photonic corrections to    
$\rho^+ \to \pi^+\pi^0$ decays.}
\end{figure}

   Just to illustrate the cancellation of infrared divergencies in the
sum of virtual and real photon corrections, we
reproduce here the expression for the virtual corrections in the case of
$\rho^0 \to \pi^+\pi^-$ decays \cite{Cremmer:1969er}:
\bea
\frac{\Gamma_0^{\rm 1v}}{\Gamma_0^0}&=& 
\frac{\alpha}{\pi}\left[\pi^{2}\left(\frac{1+v^{2}_0}{2v_0}\right)-2\left(1+
\ln\left[\frac{\lambda}{m_{\pi^+}}\right]\right)\left(  1+
\frac{1+v^2_0}{2v_0}\ln\left[\frac{1-v_0}{1+v_0}\right]\right)\right. 
\nonumber\\ &
&\left. -\left(\frac{1+v^2_0}{v_0}\right)\left[{\rm Li}_2(v_0)-
{\rm Li}_2(-v_0)\right]-\frac{1+v^2_0}{2v_0}\left(
{\rm Li}_2\left[  \frac{2}{1+v_0}\right]-{\rm
Li}_2\left[\frac{2}{1-v_0} \right]\right)\right]\ .
\label{virtualn}
\eea
The corresponding result in the case of $\rho^+ \to \pi^+\pi^0$ decays
can be computed using the procedure described above \cite{Yennie:63}.
We obtain:
\bea
\frac{\Gamma_+^{\rm 1v}}{\Gamma_+^0}&=& 
\frac{\alpha}{\pi} \left\{ -1 -2\ln \left(\frac{\lambda}{m_{\rho^+}}
 \right) \left[1+\frac{1}{2v'_+}\ln \left( \frac{1-v'_+}{1+v'_+}\right)
\right] + \frac{3}{4}\ln \left(\frac{m_{\pi^+}^2}{m_{\rho^+}^2}
\right)\right. \nonumber \\
&& \left. + \frac{m_{\rho^+}^2}{4m_{\pi^0}^2}\left[\ln
\left(\frac{m_{\rho^+}^2}{m_{\pi^+}^2}\right) \left(1-
\frac{\Delta^2_{\pi}}{m_{\rho^+}^2} \right) - v_+\left[ \ln \left(
\frac{1-v_+-\frac{\Delta_{\pi}^2}{m^2_{\rho^+}}}{1+ 
v_+-\frac{\Delta_{\pi}^2}{m^2_{\rho^+}}}\right) + \ln \left( 
\frac{1+v_+-\frac{\Sigma_{\pi}^2}{m^2_{\rho^+}}}{1-
v_+-\frac{\Sigma_{\pi}^2}{m^2_{\rho^+}}}\right)
\right] \right] \right. \nonumber \\
&& \left. +  
\frac{1}{2v'_+} \left[ \ln \left(\frac{1-v'_+}{1+v'_+}\right) \left[
-\frac{1}{4}
\ln \left(\frac{1-v'_+}{1+v'_+} \right) +2\ln\left( \frac{2v'_+}{1+v'_+}
\right) -\ln \left(\frac{m_{\rho^+}}{m_{\pi^+}} \right) \right] 
\right. \right. \nonumber \\ 
&& \left. \left. -\frac{\pi^2}{3} + 2{\rm Li}_2\left(\frac{1-v'_+}{
1+v'_+} \right) + \ln^2 \left(\frac{m_{\pi^+}}{m_{\rho^+}} \right)
\right. \right.  \nonumber \\ 
&& \left. \left. + 2{\rm Li}_2\left(\frac{1+v'_+-2/\xi}{1+v'_+}\right) +
2{\rm Li}_2\left(\frac{-1+v'_++2/\xi}{2/\xi}\right) \right] 
\right\}
 \ ,
\label{virtualc}
\eea
where we have defined $\Sigma_{\pi}^2 =m_{\pi^+}^2+m_{\pi^0}^2$. 
As pointed out before, this result is free from ultraviolet divergencies,
and it depends logarithmically on the infrared mass regulator
$\lambda$.

  As we can easily check, the sum of the virtual corrections, Eqs.
(\ref{virtualn},\ref{virtualc})  and the corresponding soft-photon
bremsstrahlung, Eqs. (\ref{bssn},\ref{bssc}), are free from the
infrared regulator $\lambda$ as it must be according to the
Bloch-Nordsiek theorem. Note however, that this sum
depends on the photon energy cut $\omega_0$. We can expect that this
$\omega_0$
dependence will be largely cancelled in the photon-inclusive $\rho \to
\pi\pi$ decay rate in such a way that $\Delta \Gamma_{\rho}$ defined in
Eq. (5) does not depended on this parameter. Since the soft-
and
hard-photon decay rates have a different dependence upon $\omega_0$, we
need to choose values of $\omega_0$ which are sufficiently small
(typically much smaller than the masses of hadronic particles)
\cite{Rodrigo:2001kf}.
In the following subsection we briefly comment about the decay rates for
hard photons.

\subsection{Hard-photon emission}

   In order to estimate the contributions to $\Delta \Gamma_{\rho}$ due to
photons of energy larger that $\omega_0$, the third and fourth terms in
Eq.
(5), we require the complete expressions  of the radiative decay
amplitudes corresponding to the Feynman diagrams in Figures 1 and 2. These
amplitudes
include also the model-dependent contributions shown in Figures 1d)-1f)
and 2d)-2e). The explicit contributions of the model-dependent amplitudes
can be found in Refs. \cite{Singer:1963ae, Toledo:2007, Lopez
Castro:2001qa}.

 When we evaluate the decay rates of radiative decays, we note that the
contributions of the model-dependent terms (intermediate states with
$\omega,\ a_1, \sigma$ and $f_0$ mesons) are typically two orders of
magnitude below the contributions due to model-independent terms
\cite{Toledo:2007, Bramon:1993yx, Lopez Castro:2001qa}. Therefore, such
model-dependent terms would affect the
width difference $\Delta \Gamma_{\rho}$ at a completely negligible level.
Note however that their effects have been included in our numerical
evaluations given below.

\section{Results and discussion}

  In this section we present the main numerical results of our
calculations. 
The values of the radiative corrections $\delta_{\rho^i}$ ($i=+,\ 0$)
defined in Eq. (11) are shown in Tables I and II for $\rho^0
\to \pi^+\pi^-(\gamma)$ and $\rho^+\to \pi^+\pi^0(\gamma)$ decays,
respectively. The
values of $\delta_{\rho^i}$ are tabulated as a function of the photon
energy cut $\omega_0$ for three different values of the $\rho^i$ meson
mass which are consistent with a small isospin breaking. As expected,
these radiative corrections depends only slightly on the specific value of
the $\rho$ meson mass. It is interesting to observe that in $\phi \to
K^+K^-$ decays (and other analogous quarkonia decays that occur close to
threshold) the radiative corrections are dominated by the coulombic
interaction in the final state \cite{Cremmer:1969er}, while in $\rho^0 \to
\pi^+\pi^-$ this is not the case.

   Table III shows the radiative decay rates of $\rho^i \to
\pi\pi\gamma$ normalized to the the corresponding tree-level rates
$\Gamma^0_i$. The tabulated values are defined as follows ($i=0,+$):
\be
\Delta_{\rho^i} \equiv \frac{\Gamma(\rho^i \to \pi\pi\gamma,
\omega\geq \omega_0)}{\Gamma^0_i}\ .
\ee
These values were calculated for photon energies larger that $\omega_0$
and for three different values of the $\rho$ meson mass. The values of
$\Delta_{\rho^i}$ are always positive and exhibit the typical decreasing
behavior as the lower
energy photon cut $\omega_0$ increases. We observe that these normalized
rates are in good agreement with the corresponding results reported 
long ago in Ref. \cite{Singer:1963ae} using slightly different values for 
the mass and width of the $\rho$ meson.

 The predicted branching fraction for the neutral radiative mode:
\be
B(\rho^0 \to \pi^+\pi^-\gamma, \omega\geq 50\ {\rm MeV}) = 11.5\times
10^{-3} \ ,
\ee
compares very well (they are in agreement within $1\sigma$) with the
experimental measurement $(9.9 \pm 1.6) \times 10^{-3}$
reported in \cite{pdg}, which was obtained for the same value of
$\omega_0$. For the same value of $\omega_0$, we obtain the isospin
breaking in the radiative modes to be (assuming $\Gamma_{\rho+}\approx
\Gamma_{\rho^0} = 150$ MeV):
\be
\Gamma(\rho^0 \to \pi^+\pi^-\gamma,\ \omega_0=50\ {\rm MeV}) -
\Gamma(\rho^+ \to \pi^+\pi^0\gamma,\ \omega_0=50\ {\rm MeV}) \approx 1.1\
{\rm MeV}\ ,
\ee
which is larger that the estimate $(0.45 \pm 0.45)$ MeV
considered in Ref. \cite{Cirigliano:2002pv} and essentially independent
of the $\rho$ meson mass.

   In Table IV we display the sum of the radiative corrections
$\delta_{\rho^i}$ and the radiative branching ratios $\Delta_{\rho^i}$
for a common value (775 MeV) of the neutral and charged $\rho$ meson mass.
According to the definitions given in the previous sections we have:
\bea
\label{sigma}\frac{\Gamma(\rho^i \to \pi\pi(\gamma),
\omega\leq\omega_0)+ \Gamma(\rho^i \to \pi\pi\gamma,
\omega\geq \omega_0)}{\Gamma^0_i} 
&=& 1+ \delta_{\rho^i} + \Delta_{\rho^i}\ \nonumber \\
&\equiv & 1+\sigma_{\rho^i}\ .
\eea
As we have pointed out in section II, we expect that the $\sigma_{\rho^i}$
correction term will be independent of the photon energy cut $\omega_0$ as
far as Eq. (\ref{sigma}) describes a photon inclusive rate and $\omega_0$
is an arbitrary reference value used to separate soft- and hard-photon
emission. In practice this cancellation is better for values of
$\omega_0$ (typically below a few MeV's, see Table IV) because radiative
corrections includes only the logarithmic dependence in $\omega_0$, while 
the hard-photon radiative rate  contains also linear and other
 $\omega_0$ dependent terms.

   Finally, we can evaluate the difference in decay widths arising from
different corrections. Based on previous definitions, we can write the
width difference of rho mesons, Eq. (5), as follows:
\be
\Delta \Gamma_{\rho} = \Gamma^0_0 \left[1+\sigma_{\rho^0} - 
\left(\frac{m_{\rho^+}v_+^3}{m_{\rho^0}v_0^3} \right)[1+ \sigma_{\rho^+}]
\right] + \Delta
\Gamma^{\rm sub} \ .
\ee
Once the isospin breaking in $\Delta m_{\rho}$ is known, the values of
$\Delta \Gamma_{\rho}$ can be easily evaluated from Tables
I-III, using the above expression. For illustrative purposes, we provide
the values of $\Delta \Gamma_{\rho}$ for two interesting cases (we have
used $\Gamma_0^0 = 150$ MeV and the value of $\sigma_{\rho^i}$ 
at $\omega_0=10$ MeV):
\be
\Delta \Gamma_{\rho} = \left\{ \begin{array}{c}
\hspace{-1.0cm} 0.86\ {\rm MeV}, \ \ \ {\rm if}\ \Delta m_{\rho} =0 \\
\  0.02\ {\rm MeV}, \ \ \ {\rm if}\ \Delta m_{\rho} =-3\ {\rm MeV} \\
\  1.70\ {\rm MeV}, \ \ \ {\rm if}\ \Delta m_{\rho} =+3\ {\rm MeV} \\
\end{array}
\right. \ .
 \ee
Note that the width difference is very sensitive to the size and sign of
$\Delta m_{\rho}$. This is very interesting because it can be useful as a
test of ($\Delta m_{\rho}({\rm exp})),\ \Delta \Gamma_{\rho}({\rm exp})$)
correlated values extracted from fits to experimental data. In particular,
the above results are in good agreement with the values 
($-2.4\pm 0.8,\ -0.2\pm 1.0$) MeV extracted from a
combined fit to $\tau$ and $e^+e^-$ data \cite{Schael:2005am}, but
the agreement is less good with the central values ($-3.1\pm 0.9,\ -2.3\pm
1.6$) MeV reported in \cite{Ghozzi:2003yn}.

\section{Conclusions}
 
   The isospin breaking in the $\rho^0-\rho^{\pm}$ system is a very
important ingredient to understand the current discrepancy in predictions
of the hadronic vacuum poplarization contributions to the muon
magnetic moment based in $\tau$ decay and $e^+e^-$ annihilation data.
  In this paper we have evaluated the difference in decay widths ($\Delta 
\Gamma_{\rho}$) of $\rho(770)$ vector mesons. We have considered in our
calculation the
isospin breaking in the exclusive decay modes of charged and neutral
$\rho$ mesons. In particular, we have carried out a calculation of the
radiative corrections to the dominant $\rho \to \pi\pi$ decay modes 
and have done a careful reevaluation of the differences in their
radiative $\rho \to \pi\pi\gamma$ decays. 

  We found that $\Delta \Gamma_{\rho}$ is sensitive to the isospin
breaking in the $\rho^0-\rho^{\pm}$ mass difference. This provides a
useful tool to test the (model-dependent) values of these parameters
extracted from experimental data. In particular we have found  that
positive
values of $\Delta \Gamma_{\rho}$ are favored by $|\Delta m_{\rho}| \leq 3$
MeV.

%%%%%%%%%%%%%%%%%%%%%%%%%%%%%%%%%%%%%%%%%%%%%%%%%%%%%%%%%%%%%%%
\acknowledgments
%%%%%%%%%%%%%%%%%%%%%%%%%%%%%%%%%%%%%%%%%%%%%%%%%%%%%%%%%%%%%%%

  The authors acknowledge financial support from Conacyt (M\'exico). They
are very grateful to Michel Davier for motivating this calculation and for
useful correspondence. Useful conversations with Augusto Garc\'\i a are
also appreciated.

%%%%%%%%%%%%%%%%%%%%%%%%%%%%%%%%%%%%%%%%%%%%%%%%%%%%%%%%%%%%%%%

% \vspace{-0.99cm}

\

\newpage

\begin{table}[t!]
\begin{center}
\begin{tabular}{| c| c c c| }
\hline\hline
& $m_{\rho^0}=772$ MeV & $m_{\rho^0}=775$ MeV & $m_{\rho^0}=778$ MeV \\
$\omega_0$ (MeV) & $\delta_{\rho^0}$ & $\delta_{\rho^0}$
&$\delta_{\rho^0}$ \\
\hline \hline
2 & -0.03670 & -0.03692& -0.03714 \\
4 & -0.02910 & -0.02930& -0.02949 \\
6 & -0.02465 & -0.02483& -0.02501 \\
8 & -0.02150 & -0.02167& -0.02183 \\
10 & -0.01905 & -0.01921 & -0.01937 \\
12 & -0.01705 & -0.01720 & -0.01736 \\
14 & -0.01536 & -0.01550 & -0.01565 \\
16 & -0.01389 & -0.01403 & -0.01477 \\
18 & -0.01260 & -0.01273 & -0.01287 \\
20 & -0.01144 & -0.01157 & -0.01170 \\
30 & -0.00697 & -0.00708 & -0.00720 \\
40 & -0.00378 & -0.00388 & -0.00399 \\
50 & -0.00130 & -0.00139 & -0.00150 \\
60 & 0.00074 & -0.00065 & 0.00056 \\
70 & 0.00249 & 0.00240 & 0.00232 \\
80 & 0.00401 & 0.00393 & 0.00384 \\
90 & 0.00536 & 0.00529 & 0.00521 \\
100 & 0.00659 & 0.00651 & 0.00643 \\
\hline
\end{tabular}
\end{center}
\vspace{-0.55cm}
\caption{ Radiative correction $\delta_{\rho^0}$ to the $\rho^0 \to
\pi^+\pi^-(\gamma)$ decay rate (see definition in Eq. (\ref{delta})) as a
function of $\omega_0$ and for three
different values of $m_{\rho^0}$.}
\vspace{-0.4cm}
\end{table}

\begin{table}[t!]
\begin{center}
\begin{tabular}{| c| c c c| }
\hline\hline
& $m_{\rho^+}=772$ MeV & $m_{\rho^+}=775$ MeV & $m_{\rho^+}=778$ MeV \\
$\omega_0$ (MeV) & $\delta_{\rho^+}$ & $\delta_{\rho^+}$
&$\delta_{\rho^+}$ \\
\hline \hline
2 & -0.01959& -0.01968& -0.01970 \\
4 & -0.01701& -0.01710& -0.01718 \\
6 & -0.01551& -0.01558& -0.01566 \\
8 & -0.01444& -0.01451& -0.01459 \\
10 & -0.01361 & -0.01368 & -0.01375 \\
12 & -0.01293 & -0.01300 & -0.01307 \\
14 & -0.01236 & -0.01242 & -0.01249 \\
16 & -0.01186 & -0.01192 & -0.01199\\
18 & -0.01142 & -0.01149 & -0.01155 \\
20 & -0.01103 & -0.01109 & -0.01115 \\
30 & -0.00953 & -0.00958 & -0.00963 \\
40 & -0.00844 & -0.00849 & -0.00854 \\
50 & -0.00761 & -0.00765 & -0.00769 \\
60 & -0.00692 & -0.00696 & -0.00700 \\
70 & -0.00633 & -0.00637 & -0.00639 \\
80 & -0.00582 & -0.00584 & -0.00589 \\
90 & -0.00536 & -0.00540 & -0.00544 \\
100 & -0.00495 & -0.00499 & -0.00502 \\
\hline
\end{tabular}
\end{center}
\label{Table2}
\vspace{-0.55cm}
\caption{ Radiative corrections $\delta_{\rho^+}$ to the $\rho^+ \to
\pi^+\pi^0(\gamma)$ decay rate (see definition in Eq. (\ref{delta})) as a
function of $\omega_0$ 
and for three different values of $m_{\rho^i}$.}
\vspace{-0.4cm}
\end{table}

\begin{table}[t!]
\begin{center}
\begin{tabular}{| c| c c |c c|c c| }  
\hline\hline
& $m_{\rho^{+,0}}=772$ MeV & & $m_{\rho^{+,0}}=775$ MeV & &
$m_{\rho^{+,0}}=778$ MeV
& \\
$\omega_0$ (MeV) & $\Delta_{\rho^+}$ & $\Delta_{\rho^0}$ &
$\Delta_{\rho^+}$
& $\Delta_{\rho^0}$ & $\Delta_{\rho^+}$ & $\Delta_{\rho^0}$ \\
\hline \hline
2 & 0.01544 & 0.04475 & 0.01553 &0.04497 &0.01561&0.04518\\
4 & 0.01290 & 0.03724 & 0.01297 &0.03742&0.01305&0.03761\\
6 & 0.01143 & 0.03288 & 0.01149 &0.03305&0.011556&0.03322\\
8 & 0.01039 & 0.02981 & 0.01045 &0.02997&0.01051&0.03013\\
10 & 0.00959 & 0.02745 & 0.00965&0.02760&0.00970&0.02775 \\
12 & 0.00894 & 0.02553 & 0.00900&0.02568&0.00905&0.02582 \\
14 & 0.00840 & 0.02393 & 0.00845&0.02406&0.00850&0.02420 \\
16 & 0.00793 & 0.02255 & 0.00798&0.02268&0.00803&0.02281\\
18 & 0.00753 & 0.02134 & 0.00758&0.02147&0.00762&0.02159\\
20 & 0.00717 & 0.02027 & 0.00721&0.02039&0.00726&0.02051 \\
30 & 0.00581 & 0.01624 & 0.00585&0.01635&0.00589&0.01645 \\
40 & 0.00488 & 0.01350 & 0.00492&0.01359&0.00495&0.01369 \\
50 & 0.00420 & 0.01146 & 0.00423&0.01155&0.00426&0.01163 \\
60 & 0.00366 & 0.00987 & 0.00369&0.00994&0.00372&0.01002 \\
70 & 0.00322 & 0.00857 & 0.00325&0.00864&0.00327&0.00871 \\
80 & 0.00286 & 0.00750 & 0.00288&0.00757&0.00291&0.00763 \\
90 & 0.00255 & 0.00659 & 0.00257&0.00665&0.00259&0.00672 \\
100 & 0.00228 & 0.00582 & 0.00230&0.00588&0.00232&0.00593 \\
\hline
\end{tabular}
\end{center}
\label{Table3}
\vspace{-0.55cm}
\caption{ Decay rates of $\rho^i \to \pi\pi\gamma$ (normalized to the
tree-level rates $\Gamma^0_i$, see definition in Eq. (16)) as a function
of the low photon energy
cut $\omega_0$ and for three different values of $m_{\rho}$.}
\vspace{-0.4cm}
\end{table}

\begin{table}
\begin{center}
\begin{tabular}{| c| c| c| c| }
\hline\hline
$\omega_0$ (MeV) & $\sigma_{\rho^0}$ & $\sigma_{\rho^+}$ &
$\sigma_{\rho^0}-\sigma_{\rho^+}$ \\
\hline \hline
2 & $\ 8.05 \times 10^{-3}$& $\ -4.15\times 10^{-3}$& $\ 12.20 \times
10^{-3}$
\\
4 & $\ 8.12 \times 10^{-3}$& $\ -4.13 \times 10^{-3}$& $\ 12.25 \times
10^{-3}$\\
6 & $\ 8.22 \times 10^{-3}$& $\ -4.09 \times 10^{-3}$& $\ 12.31 \times
10^{-3}$\\
8 & $\ 8.30 \times 10^{-3}$& $\ -4.06 \times 10^{-3}$& $\ 12.36 \times
10^{-3}$\\
10 & $\ 8.39 \times 10^{-3}$& $\ -4.03 \times 10^{-3}$&$\ 12.42 \times
10^{-3}$
\\
12 & $\ 8.48 \times 10^{-3}$& $\ -4.00 \times 10^{-3}$& $\ 12.48 \times
10^{-3}$\\
14 & $\ 8.56 \times 10^{-3}$& $\ -3.97 \times 10^{-3}$& $\ 12.53 \times
10^{-3}$\\
16 & $\ 8.65 \times 10^{-3}$& $\ -3.94 \times 10^{-3}$& $\ 12.59 \times
10^{-3}$\\ 
18 & $\ 8.74 \times 10^{-3}$& $\ -3.91 \times 10^{-3}$& $\ 12.65 \times
10^{-3}$\\   
20 & $\ 8.82 \times 10^{-3}$& $\ -3.88 \times 10^{-3}$& $\ 12.70 \times
10^{-3}$\\
\hline
\end{tabular}
\end{center}
\label{Table4}
\vspace{-0.55cm}
\caption{ Photon inclusive corrections for $\rho^i \to \pi\pi\gamma$ (see 
definition in Eq. (19)) as a
function of the photon energy $\omega_0$. The masses of charged and
neutral  $\rho$ mesons were fixed to 775 MeV's.}
\vspace{-0.4cm}
\end{table}

%%%%%%%%%%%%%%%%%%%%%%%%%%%%%%%%%%%%%%%%%%%%%%%%%%%%%%%%%%%%%%%%%%%

\end{document}